\def\year{2022}\relax
\documentclass[letterpaper]{article} 
\usepackage{aaai22}  
\usepackage{times}  
\usepackage{helvet}  
\usepackage{courier}  
\usepackage[hyphens]{url}  
\usepackage{graphicx} 
\usepackage{subcaption}
\usepackage[table,xcdraw]{xcolor}
\usepackage{makecell}
\usepackage{multirow}
\usepackage{natbib}  
\usepackage{caption} 
\usepackage{pdfpages} 
\DeclareCaptionStyle{ruled}{labelfont=normalfont,labelsep=colon,strut=off} 
\usepackage{algorithm}
\usepackage{algorithmic}

\usepackage{newfloat}
\usepackage{listings}
\floatstyle{ruled}
\newfloat{listing}{tb}{lst}{}
\floatname{listing}{Listing}
\title{More than Memes: A Multimodal Topic Modeling Approach \\ to Conspiracy Theories on Telegram}
\author{
    Elisabeth Steffen
}
\affiliations{
    TU Berlin, HTW Berlin \\
    steffen@htw-berlin.de

}

\begin{document}

\maketitle

\begin{abstract}
To address the increasing prevalence of (audio)visual data on social media, and to capture the evolving and dynamic nature of this communication, researchers have begun to explore the potential of unsupervised approaches for analyzing multimodal online content.
However, existing research often neglects visual content beyond memes and, in addition, lacks methods to compare topic models across modalities. Our study addresses these gaps by applying multimodal topic modeling for analyzing conspiracy theories in German-language Telegram channels.
We use BERTopic with CLIP for the analysis of textual and visual data in a corpus of ${\sim}40,000$ Telegram messages posted in October 2023 in 571 German-language Telegram channels known for disseminating conspiracy theories. Through this data set, we provide insight into unimodal and multimodal topic models by analyzing the symmetry and intersections of topics across modalities.
We demonstrate the variety of textual and visual content shared in the channels discovered through the topic modeling, and propose a conceptual framework for the analysis of textual and visual discursive strategies in the communication of conspiracy theories. We apply the framework in a case study of the topic group \textit{Israel Gaza}.
\end{abstract}


\section{Introduction} 
\noindent 
The rise of social media has enabled the dissemination of harmful online content, including conspiracy theories (CTs), at an unprecedented speed and scale, and has promoted the formation of communities around them \cite{bessi_users_2016,cinelli_echo_2021}. Despite the increasingly multimodal nature of social media, research on CTs has predominantly focused on textual data. According to \citet{mahl_conspiracy_2022}, 78.5\% of the studies in this area analyze text, while less than 10\% examine (audio)visual data. Unsupervised methods offer an opportunity to develop flexible approaches to analyze evolving online content. They can be leveraged to explore large amounts of unstructured data, without the need for human-labeled training corpora. A small number of recent studies explores the potential of such approaches for clustering and analyzing multimodal online communication \cite{zannettou_origins_2018,zannettou_quantitative_2019,qu_evolution_2023,theisen_n-modal_2024}.
However, existing research often neglects visual content beyond memes and, in addition, lacks methods to compare topic models across modalities. Therefore, many questions remain: What information does each modality convey? How do different modalities interact? What are the specific contributions of each modality? 
We contribute to this field of research by exploring multimodal topic modeling of text-based, image-based, and text-image-based communication on German-language conspiracist Telegram channels.\footnote{We use the term `multimodal' to refer to text-image-data, although it can encompass a variety of modalities.}  We apply BERTopic \cite{grootendorst_bertopic_2022} with the vision language model CLIP \cite{radford_learning_2021} to text, image, and text-image data. We analyze data from October 2023, a period marked by the Hamas attack on Israel and the Israel-Gaza war, which represent crisis-like events that often spur the proliferation of CTs \cite{van_prooijen_conspiracy_2017}. Using this data set, we offer insights into both unimodal and multimodal topic models, highlighting the diverse textual and visual content shared in the channels as revealed through the topic modeling process. Further, we aim to understand whether there is correspondence or overlap of similar topics across modalities.

Our work is guided by the following research questions: 
\begin{itemize}
\item RQ 1: What kinds of content (topic types) and visuals (image types) can be identified in the data set through text-based, image-based, and multimodal topic models?  

\item RQ 2: How do similar topics from different topic models correspond or intersect across modalities?

\item RQ 3: What kinds of textual and visual discursive strategies of conspiracy theories does a qualitative analysis of the results reveal?
\end{itemize}

For RQ1, our analysis reveals that most identified topics exhibit a clear thematic focus, while others fulfill functional roles, such as advertising or serving as captions. We further observe a variety of image types, with photos and screenshots as the predominant image types of topics, whereas memes play a relatively minor role. This underscores the need for a broader focus on diverse forms of visual online content, beyond memes. 
Regarding RQ2, our analysis yields one main topic, \textit{Israel Gaza}, consistent across all modalities, followed by topics related to protests and the war in Ukraine. Although similar topics appear across modalities, they are represented by different sets of documents in each modality, with text, image, and multimodal topics frequently showing limited to no overlap. Our examination of topic relations indicates less correspondence than anticipated for similar topics across modalities. Different input modalities yield divergent document sets for the same topics. While some topics share a core set of common documents across modalities, others lack such overlap. 
Our findings suggest that combining modality-specific models could enhance topic modeling by producing richer and more comprehensive topic representations compared to unimodal models. However, further research is needed to better understand the mechanisms and factors underlying the differing patterns of topics across modalities.
Extending our analysis to data from a later time period, November 2024, shows that the discovered characteristics regarding symmetry and varying intersection patterns are quite consistent, even under topic shifts. We also demonstrate the applicability of a generative Large Language Model (LLM) for automated topic annotation to improve efficiency.\\
Drawing on prior research and insights from our data set, we propose a conceptual framework for analyzing textual and visual discursive strategies in CTs (RQ 3). We apply this framework in a case study, demonstrating that integrating textual and visual discursive strategies provides a more holistic understanding of how conspiracy theories are communicated through a qualitative perspective. The case study reveals how key strategies in the dissemination of conspiracy theories can be conveyed through text and images, or both; including providing evidence, authenticating claims, discrediting mainstream narratives, exposing connections, constructing enemy images, framing ingroups and outgroups, issuing directive statements, and fostering emotional responses. To the best of our knowledge, no such framework exists yet for joint text and image analysis.

\section{Related Work} 
Visuals play an important role for the communication of CTs: images appear to provide immediate `knowledge', and are powerful means in presenting complex narratives \cite{caumanns_conspiracy_2020}. A small number of studies examine the relevance of visual content in CTs: \citet{tuters_deep_2022} analyze over 470,000 CT-related Instagram posts, guided by the concept of narrative convergence, showing that different conspiracy narratives (e.g., anti-vax, QAnon, and anti-5G), converge over time, with figures like Donald Trump and Bill Gates playing central roles in connecting them.
\citet{holzer_zwischen_2021} finds that visual communication triggers more activity than text, and is strategically used to circumvent detection mechanisms, or, as \citet{al-rawi_far-right_2024} shows, to tactically disguise hateful content as humor using emojis or memes. 
Qualitative research of CT-related memes has shown that they serve to build ingroup community and criticize outgroups like governments and elites \cite{guer_seker_conspiracymemes_2022}. However, other research on far-right visual communication on social media reveals that memes are less prevalent compared to other image types such as photographs and promotional materials \cite{mcswiney_sharing_2021}, or screenshots of news and social media posts \cite{sosa_multimodal_2022}. This indicates an important gap in existing research with its focus on memes.

Clustering and topic modeling (and combinations thereof) are widely used approaches for partitioning data into meaningful groups. Depending on the underlying mathematical approach, they not only assign documents to one or more groups, but might also allow to discover latent themes within large corpora, each representing a ``human-interpretable semantic concept'' \cite{abdelrazek_topic_2023}. 
Respective methods involve technical steps such as feature extraction, typically via pre-trained models, and the application of clustering algorithms, but also human-guided steps such as validation and interpretation of results \cite{peng_automated_2023}. While these approaches allow for the discovery of textual, visual or multimodal `topics', their application to social and political phenomena comes with the challenge of connecting automatically generated clusters with theoretically meaningful concepts post-hoc, a process which requires background knowledge and cultural understanding \cite{peng_automated_2023}. 
The choice of modality is relevant to the technical steps (i.e. how the data is processed, what features are extracted), but is also likely to influence the human interpretation of the topics.  
 
Text-based topic modeling is a common approach for analyzing CTs and other deceptive content on various social media platforms such as Twitter, Reddit, or Telegram. Researchers use it to analyze German far-right communities on Telegram \cite{wich_introducing_2021,zehring_german_2023}, the QAnon movement \cite{hoseini_globalization_2021}, CTs about COVID-19 \cite{vergani_hate_2022,kant_iterative_2022,ng_exploratory_2024}, and Russian disinformation about the Ukraine war \cite{hanley_partial_2024}. 

\subsection{Image-Based Approaches}
Unsupervised approaches are also leveraged to analyze image and text-image data on social media. Much of the existing research is on memes, and thus a specific type of visual content. The work of \citet{zannettou_origins_2018} uses image hashing and clustering to analyze the dissemination of memes on the fringe sites Gab and 4chan, and the mainstream platforms Twitter and Reddit. Their findings indicate that politics-related memes are prevalent across both types of platforms, highlighting their potential role in supporting or attacking political figures. \citet{zannettou_quantitative_2019} use the same pipeline to analyze antisemitic memes on Gab and 4chan. 
\citet{theisen_automatic_2020} develop a pipeline for analyzing meme genres through feature extraction, indexing, graph construction, and clustering. Their method reveals five `super-genres' of meme clusters, including `Political' (27\%), and `Tweet or Text' (18\%), indicating that screenshots and other images containing text represent a relevant portion of visual online communication. 

\subsection{Multimodal Approaches}
Multimodal approaches extend image-based approaches by including textual data, often in the form of image captions \cite{peng_automated_2023}. The work of \citet{li_novel_2023} jointly models text and image data with multiple labels, using a bag of visual words for images and word embeddings for text. The data set is sourced from an e-commerce platform, and therefore contains distinct visual categories, and rather short and descriptive texts. However, social media data are noisier, have different image and text types, and various text-image relations, i.e., the text is not always a description of the image content. The work of \citet{qu_evolution_2023} adapts a CLIP model to this domain by fine-tuning it on a 4chan data set. Text, image, and text-image embeddings obtained from the fine-tuned model are then clustered, finding that dominant clusters are captured in all three settings, however with a high noise level of $\sim$50-60\%. Unimodal embeddings form clusters by textual or visual similarity, while text-image embeddings cluster by common semantics. 
Our work extends these studies by examining a broader range of textual and visual content. Additionally, we apply the approach in the novel context of Telegram, a hitherto unexplored platform in this regard, and the German-language conspiracist movement. 

\section{Data and Methods}
Telegram was selected as the data source due to its important role in spreading deceptive political content, including CTs. The platform's lax content moderation, encrypted communication, unlimited message forwarding, and capacity for large broadcast channels make it appealing to political protest movements, but also extremists and violent actors \cite{baumgartner_pushshift_2020,urman_what_2022}. Previous research highlights Telegram's importance for the German far-right \cite{fielitz_hate_2020,marcks_kanalisation_2023} and for disseminating CTs \cite{hoseini_globalization_2021,vergani_hate_2022,imperati_conspiracy_2023,alvisi_unraveling_2024}. 

\subsection{Data Collection}
We collected text and image data from 683 German-language Telegram channels, previously identified as disseminators of CTs and related content by a research project.\footnote{https://dashboard.bag-gegen-hass.net/, accessed: 2023/17/08}
We relied on six CT-related categories (see Table \ref{tab:messages_per_cat}), including channels from the German \textit{Querdenken} and \textit{Reichsbürger} movements.\footnote{`Querdenken' emerged from the COVID-19 pandemic protests against government measures, uniting far-right, conspiracist, and esoteric groups. The antisemitic `Reichsbürger' movement questions Germany’s sovereignty. It was also part of the COVID-19 protests, visible through the display of the German Reich flag \cite{rathje_reichsburger_2021}.}
Data was collected using \textit{Telethon}\footnote{https://docs.telethon.dev/en/stable/} between November 4 and 6, 2023, covering all of October 2023. The initial data set contained 244,569 messages, with 15.5\% being text-only, and 38.5\% including images, highlighting the importance of analysis beyond text. 

\begin{table}
    \centering
    \begin{tabular}{|p{3.4cm}|p{2.1cm}|p{1.4cm}|}
    \hline
      \textbf{Channel Category}  & \textbf{Messages} & \textbf{Channels} \\ \hline
       Conspiracy Ideology  & 12,177 (30\%) & 186   \\
       QAnon  & 11,894 (30\%) & 73\\    
       Esotericism & 5,623 (14\%) & 89 \\
       Querdenken & 4,508 (11\%) & 125 \\
       COVID-19 Disinfo & 3,538 (9\%) & 72\\
       Reichsbürger & 2,370 (6\%) & 26 \\
       \hline
       \textbf{Total} & \textbf{40,110} & \textbf{571} \\ \hline
    \end{tabular}
    \caption{Channel category distribution in the base corpus (percentages rounded).}
    \label{tab:messages_per_cat}
\end{table}


To obtain meaningful and comparable topic models across modalities, we removed all service messages (which log channel events, such as users joining), kept only messages that contain both text and image, and applied basic text cleaning (removing URLs and user mentions, parsing hashtags) yielding 62,239 messages in total.
For each image, we calculated a dhash \cite{krawetz_kind_2013} using the \textit{imagehash} library\footnote{https://pypi.org/project/ImageHash/}. We deduplicated messages based on the cleaned text and dhash pair, yielding a base corpus of 40,110 messages, most of which were from the categories \textit{Conspiracy ideology} and \textit{QAnon} (see Table \ref{tab:messages_per_cat}).

The base corpus still contained many texts and images occurring multiple times, leading to topics dominated by these duplicates. We applied stricter, modality-specific deduplication, resulting in a text-based (36,717 messages), image-based (34,850 messages), and text-image-based (31,982 messages) subcorpus. Topic modeling was conducted on both the base and the deduplicated subcorpora.\footnote{The data sets will be made available upon request for academic use through Zenodo following FAIR principles. Due to potential legal and ethical concerns, image data will not be shared directly, but reconstruction details will be provided.}

\subsection{Topic Modeling}
We used BERTopic, which applies UMAP to embeddings for dimensionality reduction, and then clusters them with HDBSCAN. A recently added multimodal CLIP-based backend also allows to process image and text-image input.
We used the BERTopic standard multilingual model \texttt{paraphrase-multilingual-MiniLM-L12-v2} to generate embeddings for text and the \texttt{clip-ViT-B-32} model for visual and multimodal data.
Dimensionality reduction and clustering were performed using default hyperparameters. Note that the image clusters are based on the image embeddings, while the topic representations are based on image captions automatically generated by the \texttt{nlpconnect/vit-gpt2-image-captioning} model, as recommended for BERTopic. Such automated captions allow to handle image-only data sets or corpora with very short texts, which are often poorly suited for topic modeling, but quite common in social media. 
The multimodal topics are based on averaged text-image embeddings, while topic representation is derived from text only.

HDBSCAN is a soft clustering method that labels noisy data as outliers, rather than forcing them into clusters. In BERTopic, this often results in many documents left unanalyzed, a common issue with this approach \cite{de_groot_experiments_2022}. We mostly used standard recommended settings for manageability, leaving alternative clustering algorithms and hyperparameter variations for future research.
We varied the minimum number of documents assigned to a cluster (\texttt{min\_topic\_size}, values 10, 20, 50) and examined the effect of modality-specific deduplication using the base corpus and the subcorpora. Computation was performed on a server with two Nvidia A30 GPUs, an Intel Xeon Gold 6346 CPU, and 251 GB RAM. Execution times were $\sim$1 minute for text, $\sim $15-35 minutes for images, and $\sim \!$ 15 minutes for text-image data, faster than image-only due to skipping  of image captioning.\footnote{The code is available under: \url{https://github.com/eli-quereli/multimodal-tm/}}

\subsubsection{Evaluation}
Our quantitative and qualitative evaluation of different hyperparameter configurations focused on the number and size of topics, the number of outliers, and a manual inspection of topic quality based on their textual and/or visual representations. A topic was considered high quality if its focus was clearly conveyed through its name and its representative documents. Conversely, vague or thematically mixed names and documents indicated low quality.
Topic quality was similar for \texttt{min\_topic\_size} of $10$ or $20$, but $10$ led to more outliers and more topics, increasing the manual annotation effort. A \texttt{min\_topic\_size} of $50$ reduced outliers but decreased topic quality. Modality-specific deduplication improved topic quality. The optimal configuration across modalities was thus $\texttt{min\_topic\_size}=20$, with modality-specific deduplication (see Table \ref{tab:topic_modeling_results_subcorpora_optimal}). 

\begin{table}
    \centering
    \begin{tabular}{|l|l|l|l|l|l|} \hline
        \textbf{Modality} & \textbf{Messages} & \textbf{Topics} & \textbf{Outliers} \\ \hline
        Text & 36,717 & 167 & 15,189 (41.4\%) \\ 
        Image & 34,850 & 157 & 12,302 (35.3\%) \\ 
        Multimodal & 31,982 & 146 & 11,429 (35.7\%) \\ \hline
    \end{tabular}
    \caption{Topic modeling results for modality-specific deduplication and  $\texttt{min\_topic\_size} = 20$.}
\label{tab:topic_modeling_results_subcorpora_optimal}
\end{table}

\subsubsection{Annotation and Grouping} 
The validation and interpretation of topic modeling results can be challenging, and available quantitative measures do not necessarily correspond with human judgment. Manual inspection of topics and the assignment of names (topic assignment) is a common validation approach \cite{zhang_image_2022,ying_topics_2022}.
In the context of our study, topic annotations were conducted by the author, who has field knowledge and practical experience in social media data annotation. In addition to the assignment of a topic name, the topic type and image type of each topic were annotated. The corresponding label sets were developed inductively after an initial exploration of the results.

The category \textbf{topic type} characterizes the general type of a topic and helps determine its relevance for further analysis. A topic was labeled as \textit{content}, if its main thematic focus could be identified.  
In contrast, the type labels  \textit{advertising} and \textit{caption} were assigned to functional topics deemed irrelevant for content analysis. The topic type \textit{content unclear}, was used for topics with ambiguous or indeterminate thematic focus.
Each topic was assigned exactly one topic type label. \textbf{Image type} was assigned to image and text-image topics to distinguish different visual types. The initial label set included \textit{photo}, \textit{photo with text}, \textit{photo comic}, \textit{photo black-white}, \textit{screenshot news}, \textit{screenshot social media}, \textit{meme}, \textit{quote image}, \textit{graphic}, \textit{infographic}, \textit{book cover}, \textit{poster}, \textit{document}, \textit{template}, and \textit{map}. Each topic received exactly one label. Due to the limited instances in some categories, the label set was later revised to a subset (\textit{photo}, \textit{screenshot news}, \textit{screenshot social media}, \textit{meme}, \textit{infographic}, and \textit{map}), ensuring a manageable number of well-represented categories.

Text topics were labeled based on topic keywords and representative documents. Image topics were based on nine representative images selected by the model. The textual representation of image topics, based on the generated captions, was generally not useful for annotation. Multimodal topics were labeled using text and image representations. For each \textit{content} topic, a topic name was then assigned by the annotator.

Finally, we manually grouped the topics into broader topic groups by compiling them into a single spreadsheet and analyzing thematic similarities across modalities. For example, topics related to Israel or Gaza were grouped under \textit{Israel Gaza}. Topics that did not clearly relate to another topic were grouped under \textit{Misc}.

\subsubsection{Intra-Annotator Agreement}
For validation, the author re-annotated the topic type and the image type assignments on a random sample of 50 topics per modality after three days. The intra-annotator agreement, measured as Cohen's $\kappa$ (see Table \ref{tab:cohen_f1}), was moderate to high, except for the topic type in the image-only setting.
Error analysis revealed that inconsistencies in annotating image topics stemmed from relying on stylistic similarities rather than semantic coherence \citep[][cf.]{zhang_image_2022}. We resolved label conflicts to obtain the final topic and image type labels.

\begin{table}[]
    \centering
    \begin{tabular}{|l|cc|cc|cc|}
    \hline
        \multirow{2}{*}{\textbf{Category}} 
        & \multicolumn{2}{c|}{\textbf{Text}} & \multicolumn{2}{c|}{\textbf{Image}} & \multicolumn{2}{c|}{\textbf{Multimodal}} \\ \cline{2-7}
        & \textbf{ \(\kappa\)} & \textbf{F1} 
        & \textbf{ \(\kappa\)} & \textbf{F1} 
        & \textbf{ \(\kappa\)} & \textbf{F1} \\ \hline
        Topic type & 0.91 & 0.68 & 0.35 & 0.62 & 0.86 & 0.46 \\ 
        Image type & -- & -- & 0.85 & 0.74 & 0.70 & 0.51 \\ 
        \hline
    \end{tabular}
    \caption{Intra-annotator agreement (Cohen's \(\kappa\)) and LLM-based annotation performance (macro average F1) for topic types and image types.}
    \label{tab:cohen_f1}
\end{table}

\subsubsection{LLM-Based Topic Annotation}
To improve efficiency, we evaluated the use of a generative Large Language Model (LLM), \texttt{gemini-1.5-flash}, for annotating and naming topics. The model was prompted to generate concise topic names and assign topic and image types (using a simplified label set for image types) and short definitions for each label.\footnote{For prompt details please refer to the accompanying code repository.} The model's performance was evaluated against human annotations as the gold standard, with macro average F1 scores showing acceptable to good agreement (see Table \ref{tab:cohen_f1}). For topic name evaluation, we computed cosine similarity between manual and LLM-generated names using the \texttt{paraphrase-multilingual-MiniLM-L12-v2} model. The median similarity scores indicate moderate to high similarity for all modalities (text: 0.69, image:  0.62, multimodal: 0.65). The annotation process was reduced from about 4 hours (manual) to 15 minutes per modality, while topic grouping (around 30 minutes in total), remained manual to preserve human assessment.

\begin{table*}[ht]
    \centering
    {\fontsize{9}{11}\selectfont}
    \begin{tabular}{|l|r|r|r|r|r|r|}
    \hline 
     \multirow{2}{*}{\textbf{Topic Type}} & \multicolumn{2}{c|}{\textbf{Text}} & \multicolumn{2}{c|}{\textbf{Image}} & \multicolumn{2}{c|}{\textbf{Multimodal}} \\ \cline{2-7}
     & \textbf{Oct'23} & \textbf{Nov'24} & \textbf{Oct'23} & \textbf{Nov'24} & \textbf{Oct'23} & \textbf{Nov'24} \\ \hline
      Content  & 53\% & 53\% & 80\% & 89\% & 60\% & 61\% \\ 
      Content unclear & 16\% & 17\% & 17\% & 8\% & 34\% & 24\% \\ 
      Advertising & 10\% & 21\% & 3\% & 4\% & 4\% & 9\% \\ 
      Caption & 22\% & 9\% & -- & -- & 2\% & 6\% \\ \hline
      \multicolumn{7}{|l|}{\textbf{Image Type}} \\ \hline
        Photo & -- & -- & 35\% & 28\% & 43\% & 23\% \\ 
        Screenshot news & -- & -- & 11\% & 20\% & 10\% & 24\% \\ 
        Screenshot social media & -- & -- & 11\% & 23\% & 7\% & 30\% \\ 
        Infographic & -- & -- & 7\% & 5\% & 8\% & 6\% \\ 
        Map & -- & -- & 6\% & \textless 1\% & 7\% & 1\% \\ 
        Meme & -- & -- & 2\% & 5\% & 1\% & 5\% \\ 
        \hline
    \end{tabular}
    \caption{Topic and image type distribution per modality for October 2023 and November 2024.}
    \label{tab:topic_img_types}
\end{table*}

\subsubsection{Temporal Transfer}
To assess the generalizability of our insights, we analyzed data from the same Telegram channels in November 2024, covering events like the U.S. presidential elections (Nov 6th) and the breakup of Germany's `Ampel' coalition (Nov 7th). We collected 219,076 messages, 43\% with images and 13\% text-only. Using the same hyperparameters ($\texttt{min\_topic\_size}=20$, modality-specific deduplication) for topic modeling, and LLM-based topic annotation, we identified 212 topics in the text setting, and 160 topics each in the image and in the multimodal settings.

\section{Results}
\subsection{Topic and Image Types}
Examining and comparing topic types and images types across modalities allows us to answer RQ 1. As Table \ref{tab:topic_img_types} shows, the \textbf{topic type} \textit{content} is predominant across modality settings, with the highest proportion found in the image-based topic model. Interpreting image-only topics appears to be easier than text-based or multimodal topics. The higher proportion of \textit{advertising} topics in November compared to October, along with the lower proportion of \textit{caption} topics, may result from overlap in the content or characteristics of these two types. The higher proportion of \textit{content unclear} in the multimodal setting could be due to the difficulty for human annotators to interpret dual-modality topics, which requires to consider both textual and visual aspects of a topic.

The predominant \textbf{image type} category in October 2023 is \textit{photo}, followed by \textit{screenshots}, which in total account for around 20\% of the topics. The prominence of screenshots increases in November 2024, reaching around 50\%, while the photo proportion decreases but remains relevant. \textit{Memes} range between 1\% and 5\% across modalities and time periods. \textit{Maps} drop from 6-7\% in October 2023 to (less than) 1\% in November 2024, likely due to reduced relevance of geopolitical topics like \textit{Israel Gaza}. The low proportions of \textit{infographics} and maps also reflect differences in category granularity, with broader categories like photo and screenshots dominating. However, despite also being a broad category, memes remain minor, clearly indicating they are not key genres.

\subsection{Dominant Topic Groups}
The manually assigned topic groups offer a broad view of thematic differences and similarities across modalities. 
Table \ref{tab:top_five_topics} shows the five largest topic groups per modality (excluding the \textit{Misc} group due to its lack of thematic expressiveness) in October 2023. Across all modalities, \textit{Israel Gaza} dominates, reflecting the data set's focus on October 2023, including the Hamas attack on Israel (Oct 7th), and the ensuing war. Its prevalence varies by modality, with 36.2\% in text, 12.9\% in images, and 16.7\% in the  multimodal data. In the text-based setting, it forms a single large topic, while in the image and multimodal setting, it is divided into 18 and 11 smaller topics, respectively. Text captures more messages with a clear main focus, but lacks the finer granularity provided by image and multimodal settings, which offer more detailed perspectives at the cost of comprehensiveness.
The image and multimodal setting share the topic group \textit{Protests}, while \textit{Ukraine War} appears in text and image, though with differences in rank and size. The topic group \textit{Self-Love}, present in text and multimodal settings, likely originates from channels in the Esotericism channel category. Among these top-ranked topic groups, we will focus on \textit{Israel Gaza}, \textit{Protests}, and \textit{Ukraine war} in the following, as they exhibit a greater potential for CTs compared to self-love topics.

\begin{table*}[ht]
    \centering
    {\fontsize{9}{10}\selectfont}
    \begin{tabular}{|l|c|c|c|r|r|r|} \hline
       \textbf{Topic Group} & \textbf{Text Rank} & \textbf{Img Rank} & \textbf{Mm Rank} & \textbf{Text Size} & \textbf{Img Size} & \textbf{Mm Size} \\ \hline
    \textbf{Israel Gaza} & I & I & I & \textbf{4,700 (36.2\%)} & \textbf{1,900 (12.9\%)} & \textbf{2,071 (16.6\%)} \\ \hline
    \textit{Self-Love} & II &  & V & 894 (6.9\%) &  & 873 (7.0\%) \\
    \textit{Protests} &  & II & III &  & 1,666 (11.3\%) & 1,205 (9.7\%) \\
    \textit{Ukraine War} & V &  & II & 560 (4.3\%) &  & 1,491 (12.0\%) \\ \hline
    COVID-19 & III &  &  & 817 (6.3\%) &  &  \\
    German Gov &  & III &  &  & 1,251 (8.5\%) &  \\
    Conspiracy (general) & IV &  &  & 725 (5.6\%) &  &  \\
    Esotericism &  & IV &  &  & 838 (5.7\%) &  \\
    Trump &  &  & IV &  &  & 1,073 (8.6\%) \\
    Far-Right Statements &  & V &  &  & 829 (5.6\%) &  \\ \hline
    \end{tabular}
    \caption{October 2023 data set: Top five topic groups by message count, per modality. \textit{Israel Gaza} ranks first in all three settings, while three other groups (in italics), appear in the top five of two modalities. Percentages reflect modality-based proportions.}
    \label{tab:top_five_topics}
\end{table*}

In November 2024, we find three topic groups ranking in the top five across all modalities: \textit{German Politics}, \textit{Ukraine War}, and \textit{U.S. Politics}. \textit{COVID-19} appears in text and multimodal, while \textit{Finance}, \textit{Military \& War}, and \textit{Conspiracy (specific)} each rank in the top five of one modality. This marks a clear thematic shift, with \textit{Israel Gaza} losing its dominant role.

\subsection{Topic Relations Across Modality Settings}
To better understand topic relations across modality settings (RQ 2), we conducted a series of analyses examining how topics are assigned and overlap across different models.

\subsubsection{Topic Symmetry}
For each \textit{source topic}, we identified the \textit{target topic} in another modality setting with the largest intersection in terms of documents with the source topic. We then checked if the target topic maps back to the source topic, indicating symmetric alignment. The symmetry ratio, calculated as the proportion of symmetric pairs, is detailed in Table \ref{tab:topic_symmetry}. Text-image and image-text pairs have the lowest symmetry, while image-multimodal and multimodal-image have the highest, indicating stronger alignment between visual and multimodal data. The results for the November 2024 data are notably consistent. 

\begin{table}[ht]
    \centering
    {\fontsize{9}{10}\selectfont}
    \begin{tabular}{|l|l|l|} \hline  
      \textbf{Modality Pairs} & \textbf{SR Oct'23} & \textbf{SR Nov'24} \\ \hline
        Text-Img / Img-Text  & 0.15 / 0.16 & 0.17 / 0.23 \\ 
        Text-Mm / Mm-Text & 0.24 / 0.28 & 0.28 / 0.37  \\
        \textbf{Img-Mm /} \textbf{Mm-Img} & \textbf{0.58 /} \textbf{0.63} & \textbf{0.57} \textbf{0.57} \\ \hline
    \end{tabular}
    \caption{Symmetry ratio (SR) for modality pairs in October 2023 and November 2024 topics.}
    \label{tab:topic_symmetry}
\end{table}

\subsubsection{Messages Over Time} 
\label{ssec:messages_over_time} 
\begin{figure}[ht]
    \centering    \includegraphics[width=1\linewidth]{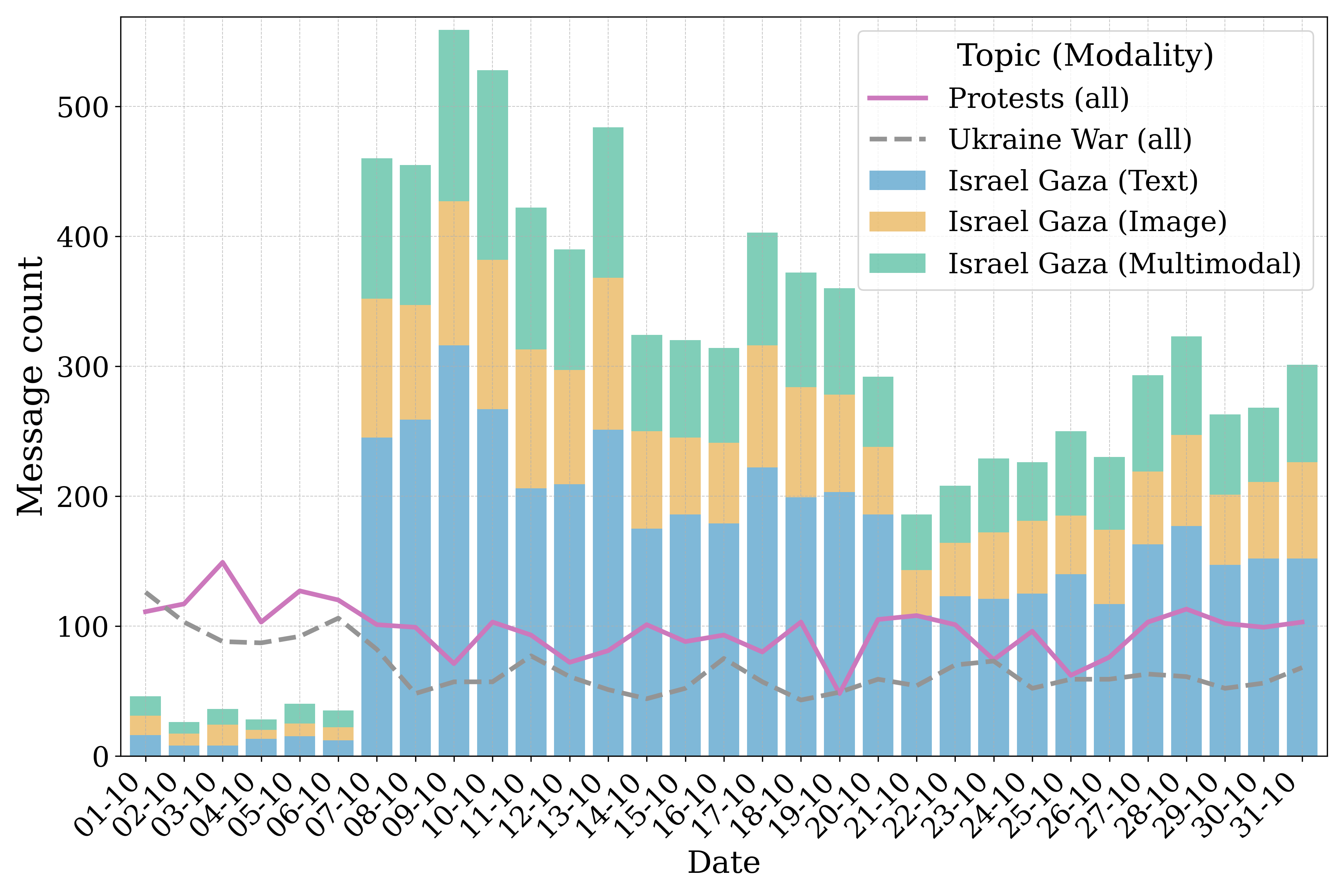}
    \caption{Message frequency over time for topic groups \textit{Israel Gaza}, \textit{Protests}, and \textit{Ukraine War}.}
    \label{fig:msg_over_time_combined}
\end{figure}

Figure \ref{fig:msg_over_time_combined} shows the message frequency over time for the dominant topic groups. \textit{Israel Gaza} shows a sharp increase on October 7th (Hamas attack on Israel) and a slight decline over time. Text dominates with about 50\% of messages, while image and multimodal make up roughly 25\% each. We found consistent modality-based proportions over time for the other (smaller) topic groups as well, with differing modality distributions. E.g., \textit{Ukraine war} is mostly represented through multimodal topics (two-thirds) and text (one-third), with marginal relevance of image topics. In contrast, \textit{Protests} is dominated by image (two-thirds), and multimodal topics (one-third), with minimal text presence.

\begin{figure}[ht]
    \centering
    \begin{minipage}{0.45\textwidth}
        \centering
        \includegraphics[width=0.8\linewidth]{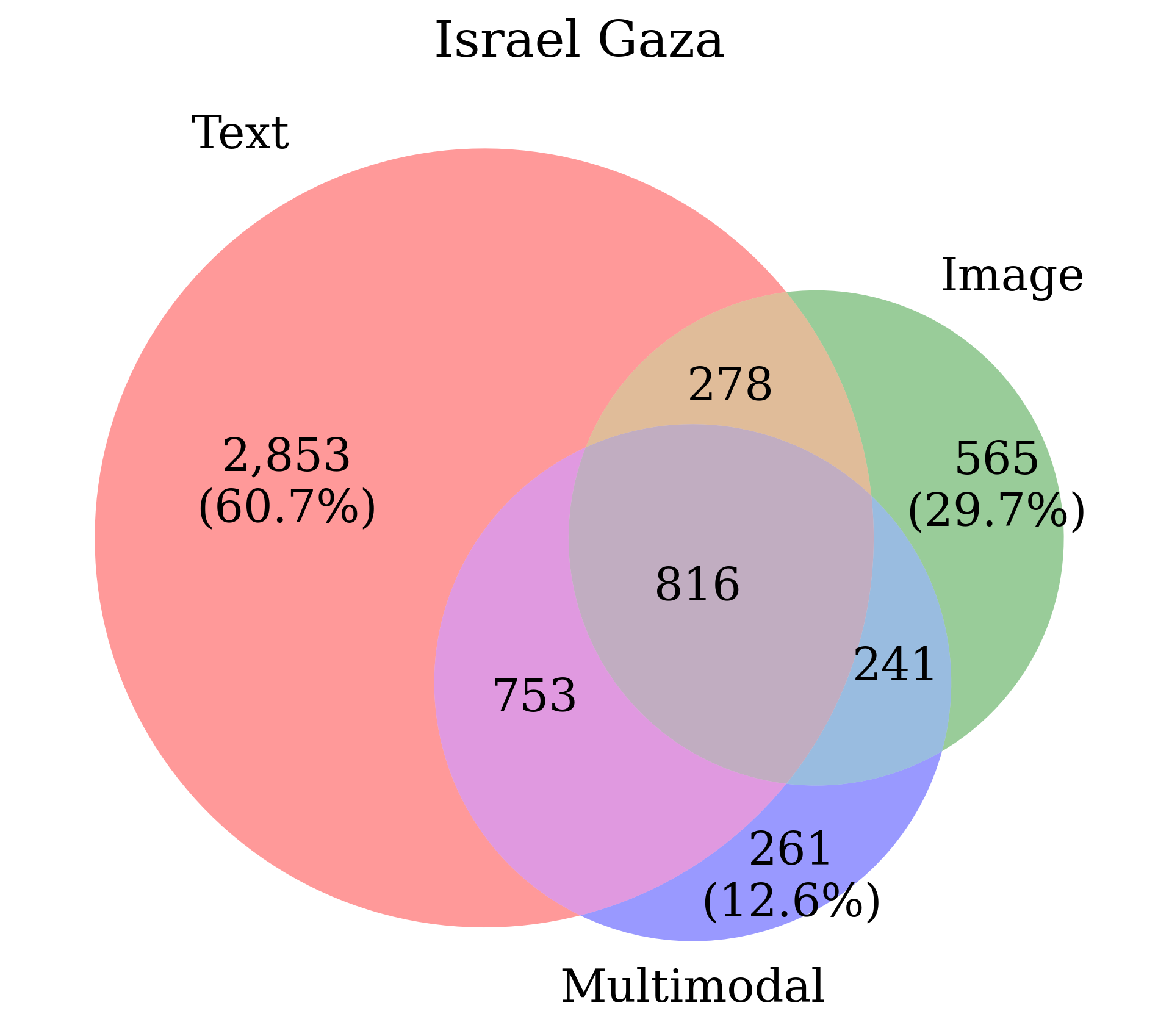}
    \end{minipage}
    \hspace{0.05\textwidth} 
    \begin{minipage}{0.45\textwidth}
        \centering
        \includegraphics[width=\linewidth]{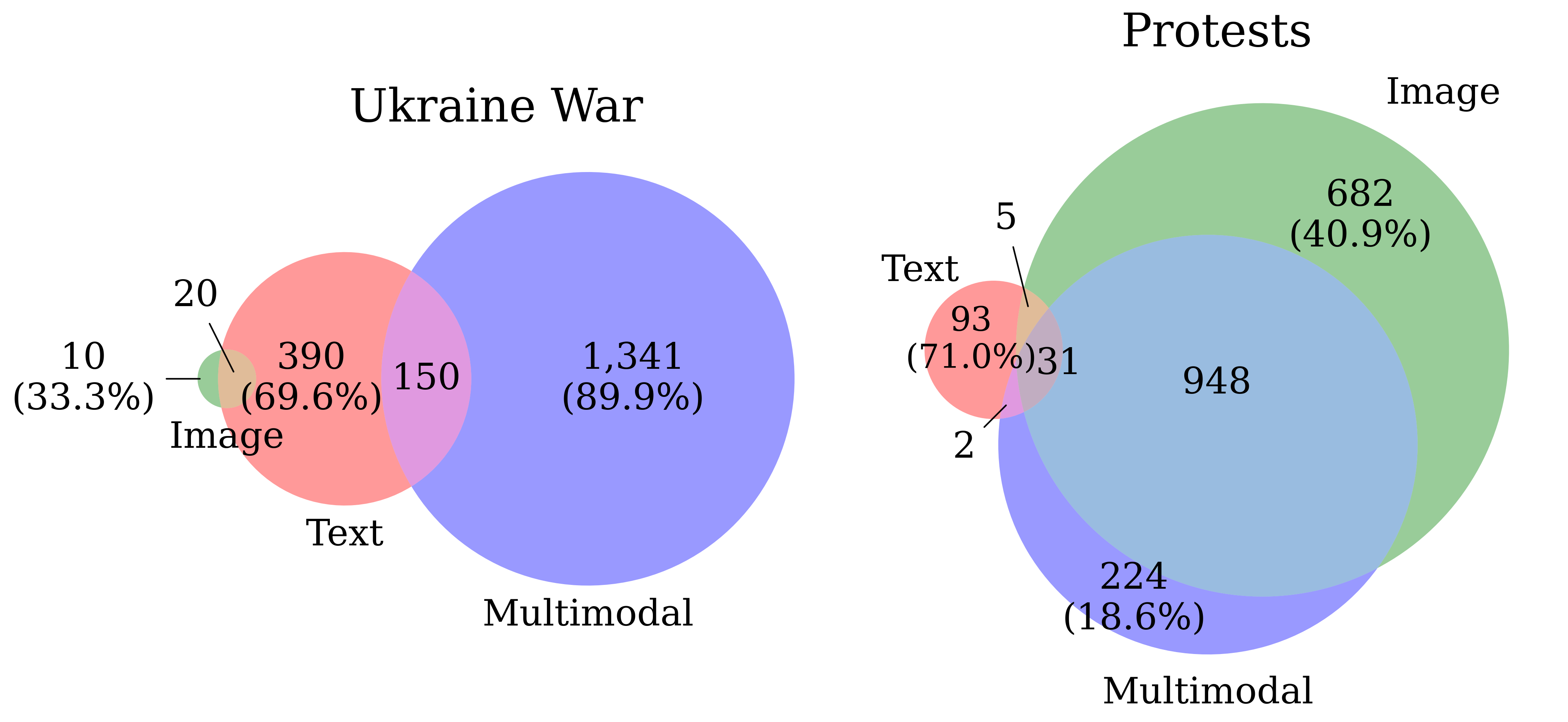}
    \end{minipage}
        \caption{Intersections of topic groups \textit{Israel Gaza}, \textit{Ukraine War} and \textit{Protests} across models. Percentages show document proportions within each modality setting.}
\label{fig:venn_israel_ukraine_protests}
\end{figure}

\subsubsection{Topic Intersections}
The topic intersections across models differ notably between the dominant topic groups in October 2023 (see Figure \ref{fig:venn_israel_ukraine_protests}). \textit{Israel Gaza} has an overlap of all three settings, but also distinct documents unique to each. Around 60\% of documents in the text group are distinct, while image and multimodal show larger overlap. In contrast, the  \textit{Protests} topic group has a very small core, but a large overlap of text and multimodal. The \textit{Ukraine War} topic group completely misses a shared core, and generally shows little overlap. Almost 90\% of the documents from the multimodal setting are not identified by the text nor the image model. While \textit{Israel Gaza} is predominantly text-based, \textit{Protests} is dominated by the image setting. For \textit{Ukraine War}, multimodal messages prevail, and the image setting is marginal. We found similar variations of overlap in the November 2024 data. These insights show that topic representation varies not only by input modality but also by topic. This requires further investigation to understand what drives modality predominance per topic, or how core documents differ from those found in only one modality setting of a topic. After this quantitative investigating of topics, we now turn to qualitative insights to explore CT-related discursive strategies in greater depth.

\section{Textual and Visual Discursive Strategies in Conspiracy Theories}
Conspiracy theories are ``efforts to explain an event or practice by referencing the machinations of powerful people'' \cite{sunstein_conspiracy_2009}. As such, they are dynamic meaning-making processes \cite{wood_propagating_2018} in which conspiracy theorists position themselves as narrators or whistleblowers, challenging established views and attempting to persuade audiences \cite{caumanns_conspiracy_2020}, and employing various \textit{discursive strategies} \cite{kou_conspiracy_2017}.

While much research examines textual strategies \citep{kou_conspiracy_2017, wood_propagating_2018}, the role of visuals remains underexplored, and no unified framework exists for analyzing both modalities. To address this gap, we propose a conceptual framework for analyzing textual and visual strategies in CTs, informed by existing research and insights from our data set. We apply this framework in a case study of the \textit{Israel Gaza} topic group to address RQ 3.
Developing this framework presents two challenges: First, it requires categories which are applicable to both textual and visual strategies. Second, the role of visuals as discursive tools can be debated due to their often ambiguous nature. However, following \citet{richardson_impact_2009}, we argue that visuals, like text, can serve as meaning-making tools, with both requiring context-sensitive interpretation.

\subsection{Conceptual Framework}
To develop a framework that captures CTs as active meaning-making processes involving both textual and visual strategies, we draw on prior research from discourse analysis in political communication, social psychology on rumors, and CTs in online discussions \cite{bordia_problem_2004, kou_conspiracy_2017, wood_propagating_2018}, narratological approaches to CT images \cite{caumanns_conspiracy_2020}, and discourse analysis of CT memes \cite{guer_seker_conspiracymemes_2022}. 
We define \textit{strategy} based on \citet{richardson_impact_2009} as:
\begin{quote}
    a (more or less accurate and more or less intentional) plan of practices, including discursive practices, adopted to achieve a particular social, political, psychological, or linguistic goal.
\end{quote}
In the context of CTs, the goal is to persuade audiences of a CT's veracity or to delegitimize official accounts \cite{caumanns_conspiracy_2020}. We thus define \textit{CT-related discursive strategies} as textual or visual techniques employed to convince audiences of a conspiracy theory's validity or to challenge and undermine official accounts. Both CT believers, aiming to `enlighten' others, and opportunistic actors, such as those using CTs in information warfare \citep{madisson_strategic_2020}, may employ these strategies. This implies that the intentionality of these strategies may vary \citep{richardson_impact_2009}, and that their use can be analyzed independently of the poster's beliefs or intentions. 

Previous studies provide important insights into textual CT-related strategies. The Rumor Interaction Analysis System (RIAS), was originally developed y \citet{bordia_problem_2004} for analyzing rumor discussions in newsgroups, categorizing strategies like authenticating claims (e.g., citing media or personal expertise), providing information, directives, and using rhetorical or sarcastic statements. Their study shows how people attempt to persuade each other, often citing personal experiences or media sources to confirm or contradict rumors. The five most common types of statements are interrogatory statements, sense-making (attempting to falsify or verify a rumor), providing information, disbelief, and digressive (unrelated) statements. The authors conceptualize rumors as efforts to reduce uncertainty in crisis-like, often ambiguous situations, typically involving events with significant impact -- characteristics shared with CTs. These commonalities motivated  \citet{wood_propagating_2018} to adapt the RIAS schema for analyzing CTs related to the Zika virus on Twitter in 2015-2016, focusing on six categories: reference, belief, disbelief, authenticating, directives, and rhetorical questions.
They find that CT-supporting tweets more often cite authorities and use rhetorical questions compared to CT-opposing posts. We incorporate authenticating, directives, and rhetorical questions into our framework, as the other categories do not represent strategies, but refer to stance, which is outside the scope of our analysis. While the work of \citet{wood_propagating_2018} offers some suitable categories, we find the range of provided strategies too limited for analyzing our data. To extend the set of strategies, the work of \citet{kou_conspiracy_2017} offers a valuable source. Their work draws on a different strand of rumor research (a study on AIDS denialism by \citet{meylakhs_aids_denialist_2014}), which they adapt to analyze CT-related posts about the Zika virus on Reddit. They identify a number of strategies in CT-supporting posts including selective citation of authoritative sources, cultural references (e.g., movies), narrative elaboration (adding unverifiable details), `connecting the dots' (linking unrelated information), ideological arguments (e.g., system critique), casting doubt, and promoting skepticism. While selective citation and connecting the dots align with authentication in \citet{wood_propagating_2018}, other strategies such as system critique, casting doubt, and promoting skepticism demonstrate how conspiracy theorists undermine official narratives while constructing alternative explanations. We consider this strategy of discrediting mainstream accounts as important aspect of CTs, and therefore include it in our framework. \\
While the studies presented so far are restricted to text data, the work of \citet{caumanns_conspiracy_2020} addresses the role of visuals in CT meaning-making. Applying a narratological approach to historical CT imagery, they identify the following key strategies: pattern recognition (e.g., network imagery), agency detection, coalition mapping (e.g., collages), enemy imagery, and visualizing secrecy. They also highlight the interplay of text and visuals in creating complex causal links or pseudo-evidence, e.g. in pseudo-scientific infographics or collages. While developed for images, we find their schema useful for our joint framework, because the presented strategies can be employed both through text and images. Finally, \citet{guer_seker_conspiracymemes_2022} highlight the role of multimodal CT posts in community building and expressing skepticism toward elites, governments, and the media. While their analysis extends beyond the post itself to include user interactions (e.g., comments, likes, profiles), which is beyond the scope of our study, their work offers valuable insights for our study: the expression of skepticism aligns with the strategy of discrediting mainstream accounts, as noted by \citet{kou_conspiracy_2017}. Additionally, the focus on community building introduces the crucial aspect of constructing an ingroup vs. outgroup dynamic in CTs. 

Building on these approaches and insights from our data set, we select categories that are applicable to both textual and visual dimensions, and suitable for our data. The resulting framework, presented in Table \ref{tab:conceptual_framework}, includes eight discursive strategies. While tailored to this study, the categories are designed for broader use with CT-related data across contexts and time periods.
\begin{table*}
    \centering
    \begin{tabular}{|p{3.1cm}|p{6.5cm}|p{6.7cm}|}
        \hline
        \textbf{Discursive strategy} & \textbf{Textual manifestation} & \textbf{Visual manifestation} \\\hline
        Providing evidence & Presenting information as factual claims or evidence & Images conveying factual authority (e.g., maps, infographics, documentary-style photos) \\ \hline
        Authenticating & Referring to perceived experts or insiders to build credibility & Quote-based sharepics or screenshots showing statements by recognized figures\\ \hline
        Discrediting mainstream accounts & 
        Employing irony, sarcasm, and derogatory language; highlighting perceived contradictions or inconsistencies; rhetorical questions & Political memes or cartoons; annotated screenshots emphasizing perceived contradictions \\ \hline
        Acts of disclosure and revealing connections & Identifying actors and alleging hidden connections, often tied to a shared malevolent agenda & Depicting actors jointly to illustrate alleged connections \\ \hline
        Constructing enemy images & Employing derogatory or dehumanizing language
    & Visual representations such as portraits, caricatures, or memes \\ \hline
        Ingroup vs. outgroup & Framing the ingroups as enlightened, rebels, or persecuted victims & Images of supportive crowds or prominent leaders or heroes (e.g., Trump, Musk) \\ \hline
        Directives & Encouraging audience engagement through directives (e.g., click a link, explore further) & Sharepics or posters designed to mobilize action (e.g., attend a demonstrations, listen to a radio show) \\ \hline
        Emotionalization & Leveraging emotionally charged language to evoke fear, anger, or other strong reactions & Employing emotionally evocative imagery to provoke fear, anger, or similar responses \\ \hline
    \end{tabular}
    \caption{Conceptual framework for the analysis of textual and visual CT strategies}
    \label{tab:conceptual_framework}
\end{table*}

\subsection{Case Study: The Topic Group \textit{Israel Gaza}}
We used the proposed framework to analyze the predominant topic group in the October 2023 data set, \textit{Israel Gaza}. For each modality, we explore the thematic aspects of the topics and analyze CT-related textual and visual discursive strategies.\footnote{Note that all text examples are presented as English translations of the posts.}

Due to the large volume of 4,700 messages in the \textbf{text-based topic}, we performed a second round of topic modeling to obtain more granular topics. Using similar hyperparameters for consistency ($\texttt{min\_topic\_size}=15$), we found 51 topics. These were manually grouped by thematic similarity, following the same process as with the initial topic modeling.  \\
The identified topics mention political and military actors (individuals and groups) in the region, and cover different aspects of military actions in Gaza, with the attack on a hospital in Gaza standing out as a distinct event. They also refer to protests against Israel's actions, and discuss the role of the U.S. administration and the German government in the war. Other topics address the history of the Israeli-Palestinian conflict, or claim to reveal the reputed `truth' about the Hamas attack on Israel. More general topics include Zionism, media propaganda, and the purported relation between Islam, immigration, and antisemitism.

Applying the conceptual framework reveals a number of discursive strategies: Topics discussing the history of the Israeli-Palestinian conflict as well as claims to `reveal the truth' about the Hamas attack can be interpreted as providing evidence. A closely related strategy are acts of disclosure and revealing connections: For instance, the topic \textit{Truth Hamas attack} suggests that the attack was actually orchestrated by Israel, the USA, `Zionists', and Prime Minister Netanyahu in collaboration with a `globalist elite'. This claim explicitly connects multiple actors and aligns them with a shared, malicious goal, reinforcing a conspiratorial worldview. 
One particularly explicit post claims that the incident was \textit{``an `inside job' by the globalist elite''} in cooperation with the USA and Israel, \textit{``as part of the grand master plan for World War III.''}
The strategy of authentication occurs in posts which claim to refer to insider knowledge (without naming them explicitly). For example, a user refers to unnamed \textit{``insiders''} and \textit{``officials''}, which creates an impression of credibility. 
Irony and sarcasm are employed to discredit mainstream accounts: The following message sarcastically criticizes the framing of oppositional views in mainstream discourse:
\textit{``Corona deniers, anti-vaccinationists and Putin haters are out! Now there are ISRAEL HATERS.''}
The statement also reflects the ingroup vs. outgroup strategy, portraying the ingroup as marginalized rebels unjustly dismissed by the mainstream.
Another post adapts Reichsbürger rhetoric to a new context by claiming that Israel is not a nation-state but a corporate entity. This aligns with the strategy of constructing enemy images by using delegitimizing language. An example of emotionalization is this post, which uses capslock and exaggerations while also serving as an act of disclosure, claiming the true goal behind the Israel-Gaza war is:
\begin{quote}
    THAT ISRAEL AND THE ZIONISTS THEN OWN THE GAZA STRIP'S TENS OF TRILLIONS OF DOLLARS WORTH OF GAS RESERVES FOR THE NEXT HUNDRED YEARS OR MORE.
\end{quote}
or by asking the audience if they want to \textit{``understand who is behind the Israel-Palestine war and what goals are being pursued?''}, prompting them to click on link for answers, attempting to involve readers further.

\begin{figure}[ht]
    \centering
    \begin{tabular}{@{}c@{}}
        \includegraphics[width=0.95\linewidth]{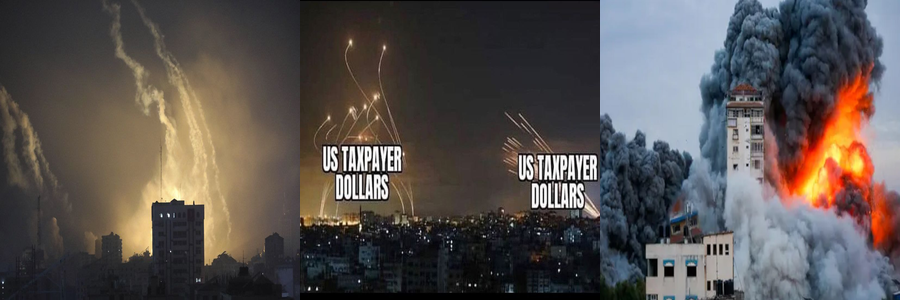}
    \end{tabular}
    \begin{tabular}{@{}c@{}}
        \includegraphics[width=0.95\linewidth]{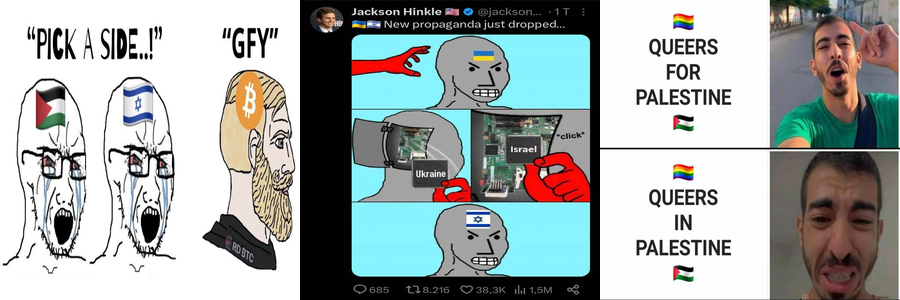}
        \label{fig:memes}
    \end{tabular}
     \begin{tabular}{@{}c@{}}
     \includegraphics[width=0.95\linewidth]{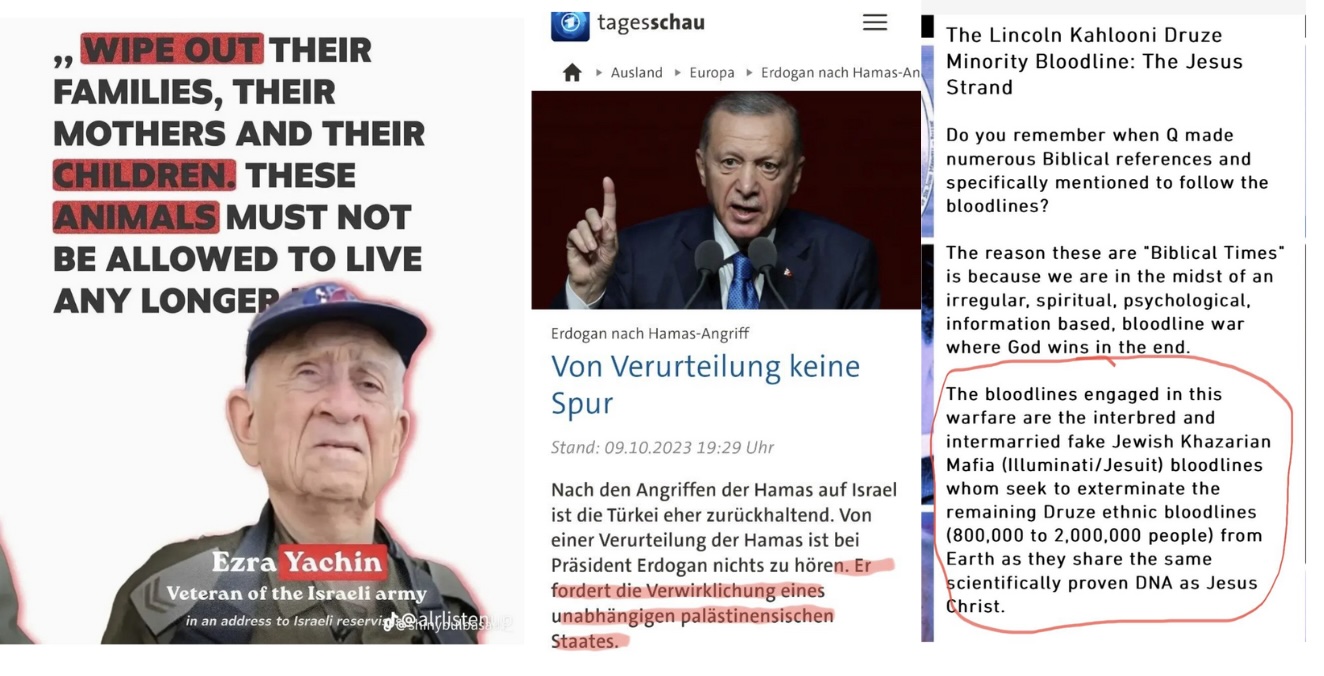}
    \end{tabular}
         \begin{tabular}{@{}c@{}}
    \includegraphics[width=0.95\linewidth]{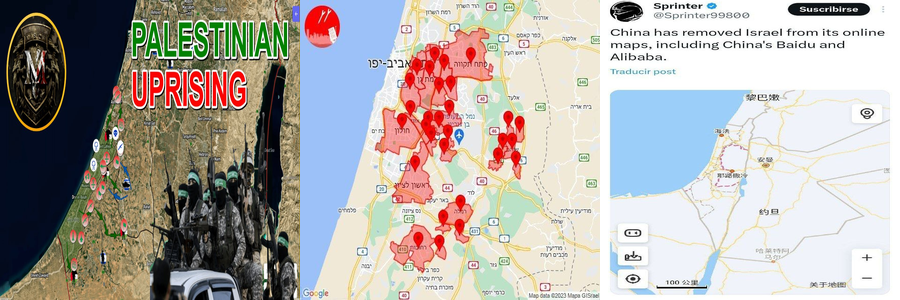}
       \end{tabular}
       \caption{Image-based model: Examples from the \textit{Israel Gaza} topic group.}
       \label{fig:image_topics}
\end{figure}

The \textbf{image-based topics} feature diverse visuals, including photos of explosions and debris, portraits of key figures (e.g., Benjamin Netanyahu), screenshots, memes, graphic representations like maps, cartoons, and symbolic images such as flags.
Like in the text setting, topics cover political and military actors, aspects of the Israel Gaza war, and the history of the conflict. Photos of explosions and debris (Figure \ref{fig:image_topics}, row 1), highlighting the war's devastation and evoking fear, exemplify emotionalization. Symbols like flags (e.g., of Ukraine and Israel, Figure \ref{fig:image_topics}, row 2) visually reveal connections, linking nations or ideologies to construct associations. Memes and cartoons (Figure \ref{fig:image_topics}, row 2) support the strategy of opposing or discrediting mainstream accounts. These visuals often employ irony or sarcasm and create links between unrelated events such as the Israel-Gaza war, the Ukraine war, and COVID-19. In doing so, they reinforce oppositional narratives through humor and simplification.
Textual elements play a central role in many images, with screenshots being the most text-heavy examples. The technique of `quoting' via screenshots, where text from other sources is directly captured and presented, aligns with strategies of providing evidence and authenticating. Highlighting details in screenshots serves to emphasize key information, expose inconsistencies, or delegitimize specific actors (Figure \ref{fig:image_topics}, row 3). Maps and infographics often serve the purpose of providing evidence but can also reflect the ideological framing. For example, some maps label the Hamas attack as \textit{``Palestinian uprising''}, implicitly framing Israel as a suppressor and supporting the construction of enemy images (Figure \ref{fig:image_topics}, row 4).

\begin{figure}[ht]
    \centering
    \begin{tabular}{@{}c@{}}
    \includegraphics[width=0.95\linewidth]{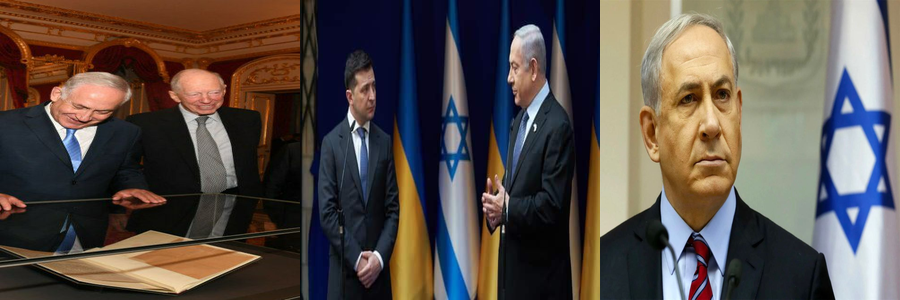}
    \end{tabular}
    \begin{tabular}{@{}c@{}}
        \includegraphics[width=0.95\linewidth]{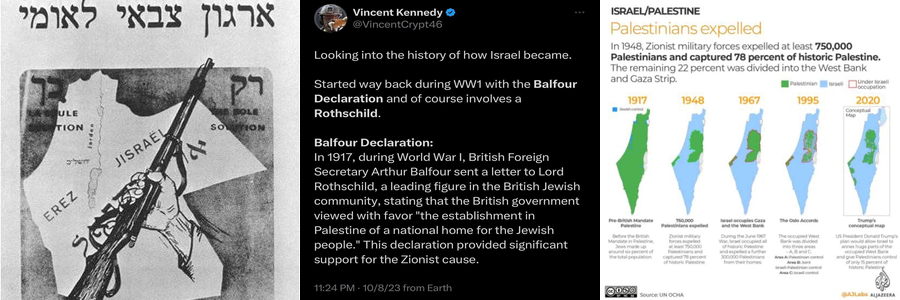}
    \end{tabular}
     \begin{tabular}{@{}c@{}}
    \end{tabular}
         \begin{tabular}{@{}c@{}}
 \includegraphics[width=0.95\linewidth]{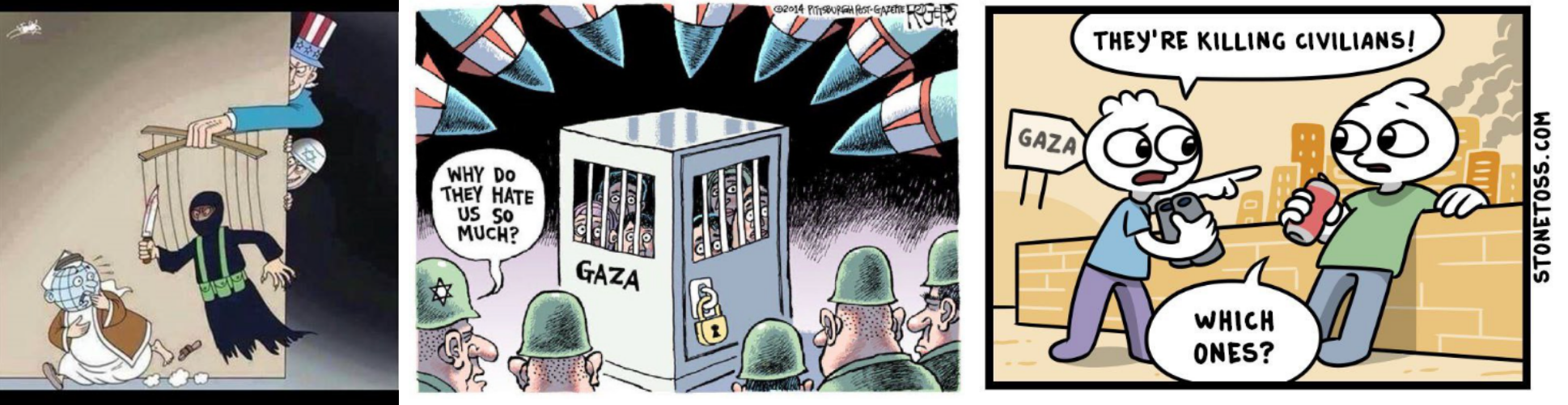}   
 \end{tabular}
       \caption{Multimodal model: Examples from the \textit{Israel Gaza} topic group.}
       \label{fig:mm_topics}
\end{figure}

The \textbf{multimodal topics} offers fewer aspects than the image setting, but the inclusion of textual information enhances interpretation and yields visually more diverse topics. We found one topic focused on actors, while others address the Israel Gaza war, protests, and geopolitical issues, including references to historical events like the topic \textit{1948 Nakba}. Figure \ref{fig:mm_topics} (row 1) shows images of Israel's Prime Minister Benjamin Netanyahu. Accompanying texts portray him as a conspiratorial actor, often in antisemitic terms, illustrating the construction of enemy images and revealing connections strategies. One post labels him a \textit{``Zionist prostitute [...] with the Zionist master lord Jacob Rothschild''}, while another accuses Netanyahu and Zelensky of being \textit{``Jewish N...s''} (omission in original) opposed to peace, and a third represents the strategy of providing evidence, referring to \textit{``internal sources''}, claiming that Netanyahu supports using force against Iran's nuclear program.
Figure \ref{fig:mm_topics} (row 2) shows examples of posts employing providing evidence and acts of disclosure: the first image, a black-and-white map purportedly from the radical Israeli group Irgun, is presented as evidence of Israel's alleged colonial agenda, with accompanying text claiming that it \textit{``is not an antisemitic invention that Israel is striving for a `Greater Israel'''}. 
The second image features a long and cryptic QAnon-style text with phrases like \textit{``1st Map Palestine 1905''} and \textit{``Balfour Declaration''}. Notably, the post was assigned as outlier in the image setting, but assigned to the \textit{Israel Gaza} topic in both the text and multimodal setting. Despite no visual similarity to the other two images in the row, apparently it's textual dimension establishes a thematic link. This post exemplifies the authenticating strategy, presenting a screenshot of a post from the platform X by a conspiracy theorist with over 400K followers, an apparently influential figure.
The third image is an Al Jazeera infographic, accompanied by text suggesting Gaza’s size could have tripled by 2020, possibly referencing then former US-American President Donald Trump's alleged plan for a larger Palestinian state on the Sinai Peninsula. The use of an infographic conveys information in a factual manner, aimed at providing evidence.

Cartoons in the context of the topic are employed for acts of disclosure, but also serve to create enemy images (Figure \ref{fig:mm_topics}, row 3). The first cartoon shows a US-American puppet master with a Jewish ally, identified by national, religious, and political symbols. The text claims Hamas and IS were founded by the USA and Jews, who are further demonized through Nazi associations. The middle cartoon depicts Gaza as a prison under constant threat of bombs, while a Jewish soldier, marked by a Star of David on his helmet, portrays Israel, represented by its army, as victim. The cartoon claims to reveal Israel's alleged evil and hypocrisy in falsely presenting itself as a victim. The accompanying text, \textit{``I love it''}, reflects the poster's support for the cartoon's message.  
The cartoon on the right conveys a similar idea, highlighting double standards in how the war is perceived: it suggests that Palestinian civilian deaths are disregarded, while Israeli deaths receive more attention. The post's text reinforces this, stating that \textit{``In the Anglo-Saxonian media it's simple: in Israel there are many murders, and in Gaza there are many deaths''}. This implies that `the media' denies the deliberate killing of Palestinians while emphasizing Israeli casualties. The cartoon thus serves the strategies of discrediting mainstream accounts, accompanied by emotionalization.

\section{Conclusion}
Our study applies multimodal topic modeling to German-language conspiracist Telegram data from October 2023, with additional analysis of a November 2024 data set to evaluate temporal transferability. We demonstrate the utility of BERTopic for analyzing German-language text, image, and text-image Telegram posts, and integrate LLM-based topic annotation to increase efficiency and practical utility of the approach. 
Regarding predominant topics, we find that in October 2023, the \textit{Israel Gaza} topic dominates across all topic models, while November 2024 focuses on \textit{German Politics}, \textit{Ukraine War}, and \textit{U.S. Politics}. Our analysis further reveals various image types, with photos and screenshots dominating, while memes play a minor role. The discovered visual types illustrate the diversity of visual content on Telegram, and also reflect varying text-image ratios: photos prioritize visuals, screenshots emphasize text, and memes represent an intermediary type.
Our analysis of topic characteristics in different models reveals that topics are communicated differently: For instance, the largest part of the \textit{Israel Gaza} topic group is present in text, while \textit{Protests} rely heavily on images. Further, each model captures a substantial number of documents not captured by the other models, resulting in often small topic intersections. Topic symmetry is lower between text and image, and higher between image and multimodal models. 
Temporal transfer overall confirms our findings regarding the distribution of topic types, image types, symmetry and intersections of similar topics across models. 
Further, building on prior research on CT-related textual and visual meaning-making, we propose a novel framework for analyzing CT-related discursive strategies, comprising eight categories suited for the joint analysis of text and image data. Applying it to the \textit{Israel Gaza} topic, we identify strategies such as providing evidence, authenticating, discrediting mainstream accounts, revealing connections, constructing enemy images, ingroup-outgroup framing, directives, and emotionalization, in text and image data.

\section{Discussion}
Our study is subject to the following limitations: first, while we selected October 2023 to capture a period marked by heightened CT activity due to major crisis events such as the Hamas attack on Israel and the Israel-Gaza war, this choice may have biased the data set toward these events. To address this, we analyzed data from a later period, confirming findings such as screenshot predominance and low symmetry, even with thematic shifts. Second, our embedding model, pre-trained on descriptive image captions, is not optimized for noisy, user-generated texts with complex text-image relationships (e.g., commentative text-image pairs) \citep[][cf.]{theisen_c-clip_2023}. Fine-tuning CLIP on Telegram data and refining multimodal embedding strategies, such as adjusting text embedding weights, could improve topic quality. Third, the absence of standardized evaluation metrics for BERTopic models introduces subjectivity and limits comparability. Expanding the annotator pool in future studies could enhance validation reliability, and provide an even more solid gold standard for the evaluation of LLM-based performance. \\ Despite its limitations, our work makes important contributions to the advancement of multimodal topic modeling. It demonstrates the potential of combining text-based, image-based, and multimodal topic models: each model identifies unique sets of documents for shared topics, and their integration can provide a more comprehensive understanding of the data set while reducing the prevalence of outlier documents. The suggested approach is particularly valuable for researchers and practitioners in social media monitoring, as it uncovers emerging themes, including image-driven topics, in dynamic contexts. Our integration of LLM-based topic annotation enhances the practical utility of our approach by increasing its efficiency. By incorporating images, our study reveals topics that are primarily expressed visually, and therefore would be missed by text-based topic models. 
We contribute to multimodal research by highlighting the variety of visual types in Telegram channels, particularly the overlooked role of screenshots. The identified image clusters can also be useful in strategic sampling of training data for supervised classification models. Our conceptual framework enhances the integrated analysis of textual and visual data, and our case study demonstrates how these strategies empirically manifest in CT-related online content.\\
Future work could explore the reasons behind modality-specific differences in topic representation. While our study highlights notable variations in topic symmetry and intersections, a deeper investigation into the underlying causes is needed. Investigating the temporal dynamics of CT content across different periods, especially during less eventful times, could provide a broader understanding of how topics evolve over time. A systematic comparison of different platforms and content types could also offer valuable insights into the cross-platform dynamics of multimodal discourse. Insights into the relevance of screenshots open new avenues for studying sharing practices, information propagation and content dissemination, as well as the role of (manipulated) screenshots in disinformation on social media. Future work could explore these practices further in order to complement existing research by including visual content. 
Our conceptual framework offers a foundation for both qualitative studies, such as multimodal discourse analysis, and quantitative studies, including annotation schema design for content analysis. It could be applied to diverse corpora, enhancing analyses of how CTs are communicated across platforms through text, images, or their combination.

\section{Ethical Statement}
Our research uses publicly available Telegram data, which may include hateful speech. While obtaining informed consent from all users is not feasible, we adhere to ethical guidelines \cite{rivers_ethical_2014} by presenting results in aggregate form, and removing personally identifiable information. The data set will be released under restricted access for academic purposes only. Potential negative effects include the misuse of data for spreading and automatically generating harmful content, and affecting individuals targeted by hateful speech. Our goal is to strengthen the potential positive impact of our work, namely improving our understanding of harmful online communication to aid in developing effective monitoring and countermeasures.

\fontsize{9pt}{10pt} \selectfont 
\bibliography{references.bib}


\end{document}